\documentclass[%
 aip,
 jcp,%
 amsmath,amssymb,
 reprint,%
]{revtex4-1}

\usepackage{subfigure}
\usepackage{graphicx}
\usepackage{dcolumn}
\usepackage{bm}

\begin{document}


\title
{Scaling properties of first-passage quantities on the fractal and transfractal  scale free networks }%


\author{Junhao Peng}
\email{pengjh@gzhu.edu.cn}
\affiliation {School of Math and Information Science, Guangzhou University, Guangzhou 510006, China.}

\affiliation {Key Laboratory of Mathematics and Interdisciplinary Sciences of Guangdong
Higher Education Institutes, Guangzhou University, Guangzhou 510006, China}

\date{\today}

\begin{abstract}
Fractal (or transfractal) and scale free characters are common properties of real life network systems. It  is of great significance to  uncover the effect of these characters on the dynamic processes taking place on complex medias. In this paper, we consider the random walk process on a kind of fractal (or transfractal) scale free networks, which also called as  $(u,v)$ flowers, and we focus on the global first passage time (GFPT) and first return time (FRT). Here, we present method to derive exactly the probability generation function, mean and variance of the GFPT and FRT for a given hub (i.e., node with the highest degree) and then the scaling properties of the  mean and the variance of the GFPT and FRT are disclosed.
Our results show that, for the case of $u>1$, while the networks are fractals,  the mean  of the GFPT scales with the volume  of the network  as $\langle GMFPT_t\rangle \sim N_t^{{2}/{d_s}}$, where  $\langle * \rangle$ denotes the mean of random variable $*$, $N_t$ is the volume of the network with generation $t$ and $d_s$ is the spectral dimension  of the network; but, for the case of $u=1$, while the networks are nonfractals, the mean  of the GFPT scales  as $\langle GMFPT_t\rangle \sim N_t^{{2}/{\tilde{d}_s}}$, where $\tilde{d}_s$ is the transspectral dimension of the network,  which is introduced in this paper.
Results also show that, the variance of the GFPT scales as $Var(GFPT_t) \sim  \langle GFPT_{t}\rangle^2$, where   $Var(*)$ denotes variance of of random variable $*$; whereas the variance of the FRT scales as $Var(FRT_t) \sim \langle FRT_{t}  \rangle \langle GFPT_{t}\rangle$.
Our results imply that for the case that the networks are nonfractals, the mean and the variance of the GFPT are not controlled by the spectral dimension(i.e., $d_s=2$), but they are controlled by the transspectral dimension.
In order to evaluate the  fluctuation of the  GFPT and FRT, we also  calculate  the reduced moments of the the GFPT and FRT and  find that, in the limit of large size, the reduced moment of the FRT tends to be infinite, whereas the reduced moment of the GFPT is almost a const.
 Therefore,  on the $(u,v)$ flowers of large size, the fluctuation of the FRT is  huge, whereas  the fluctuation of the GFPT  is much smaller.

\end{abstract}




\maketitle


\section{Introduction}
\label{intro}

 It is well known that many real-life networks display scale-free character~\cite{Albert02} and this character has profound impact on dynamics taking place on scale-free networks~\cite{ZhangGao10, TeBeVo09, AgBu09, AgBuMa10}. Nevertheless, scale-free behavior cannot reflect all the structural information of real networks. It was acknowledged that several naturally occurring scale-free networks exhibit fractal or transfractal scaling~\cite{Song05, Song06}. Taking into account fractal (or  transfractal) scaling of scale-free networks can lead to a better understanding of how the underlying systems work~\cite{Song06, zhangXie09a, zhangWu11}. However, It is difficult to determine exactly the effect of the fractal and scale-free characters on the dynamics on the real networks. In 2007, Rozenfeld \emph{et al} introduced a kind of  scale free networks which have fractal (or transfractal) character. These networks are controlled by two integral parameters (i.e., $1\leq u \leq v$) and they are also called as  $(u,v)$ flowers.  For  $u>1$, they are \textquotedblleft large-world\textquotedblright and fractals. For $u=1$, they are small-world and nonfractals, but they are  referred as transfractals~\cite{RoHa07}. These networks are of great importance because many dynamic properties can be exactly determined   and then one can  uncover the effect of the scale-free and  fractal characters on the dynamical processes taking place on them by  assigning $u$ and $v$ with different values~\cite{ZhangXie09, MeAgBeVo12, Hwang10, Zhang11}.
  The pseudofractal scale-free web~\cite{Dorogovtsev02}, which has attracted lots of interest in the past several years~\cite{zhang07, zhang07b, zhang10, ZhangLin15, ZhQiZh09, peng15}, is just an example of $(u,v)$ flower with $u=1$ and $v=2$.

Here we focus on  random walks, which is a fundamental dynamic process taking place on complex medias~\cite{HaBe87, Avraham_Havlin04, ChPe13}, on the $(u,v)$ flowers. The quantities we are interested in are  the  first passage time (FPT), which is the time it takes a random walker  to reach a given site for the first time, and first return time (FRT), which is the time it takes  a random walker to return to the starting site for the first time ~\cite{Redner07,  MeyChe11, Condamin05, CondaBe07, BeChKl10, EiKaBu07,MoDa09,SaKa08}. In the past several years, a lot of work was devoted to analyze  the mean~\cite{BeTuKo10,BlumJur03, LiZh13, Peng14d, ZhLi15, Peng14a, AgCasCatSar15,  MeAgBeVo12, CoMi10, ZhDongSheng15, LO93}  and the  variance~\cite{Redner07, KahnRed89, KoBl90,ArAnKo88, HaRo08} of the two random variables on different networks. The mean  provides  valuable estimate of the random variable and the  variance  is  good measure on whether the estimate provided by the mean  is reliable. There are also some work which discloses the relation of between spectral dimension $d_s$ and the mean and variance of FPT. For example, Haynes \emph{et al} found that on some special fractal lattices, the mean of FPT scale with the volume of network as $N^{2/d_s}$ and the variance  of FPT scale with the volume of network as $N^{4/d_s}$~\cite{HaRo08}. Hwang \emph{et al} found that the mean of FPT on scale free networks are affected by the spectral dimension $d_s$ and the exponent of the degree distribution~\cite{HvKa12}.

For random walks on the $(u,v)$ flowers, the mean of FPT (MFPT) to a given hub (i.e. nodes with the highest degree) and the average of the MFPTs from all possible starting nodes are obtained for some special $u$ and $v$~\cite{ZhangXie09, MeAgBeVo12}. The spectral dimension of the $(u,v)$ flowers with shortcuts are founded~\cite{Hwang10} and the relation between fractal dimension  and the MFPT on special kind of $(u,v)$ flowers is also analyzed~\cite{Zhang11}. However, the exactly results of the mean of FPT for any parameters $u$ and $v$, the variance of the FPT and FRT, the relations among the  mean and variance of FPT and FRT, the relation between the spectral dimension and the  mean and variance of FPT and FRT, are still all unknown.

In this paper, we study the mean and variance of FPT and FRT on the general $(u,v)$ flowers. Note that the FRT and FPT are deeply affected by the source or target site. We don't intend to enumerate all the possible cases and analyze them. On the contrary, we only consider the FRT for a given hub and the  global FPT (GFPT) to a given hub, which is  the average of the FPTs for arriving at a given hub from any possible starting site, with  the probability that the random walker starts from  node $v$ is $d_v/(2E_t)$,  where $E_t$ is the total numbers of edges of the network and $d_v$ is the degree of node $v$. Here  we  present  method to calculate exactly the mean and the variance of the GFPT  and FRT for a given hub and disclose the factors which affect them.  Firstly, we analyze and obtain the recurrence relations of the probability generating functions (PGF) of the GFPT and FRT. Then, exploiting the probability generating function tool, we  obtain the recurrence relations of the  first and second moment of the GFPT and FRT. Finally, we obtain exactly the mean and variance of the GFPT and FRT by solving the recurrence relations.  Results show that  $Var(GFPT_t) \sim  \langle GFPT_{t}\rangle^2$ and $Var(FRT_t) \sim \langle FRT_{t}  \rangle \langle GFPT_{t}\rangle$, where $Var(GFPT_t)$ and $Var(FRT_t)$ denote the variance  of the GFPT and the FRT,  $\langle GFPT_{t}\rangle$ and  $\langle FRT_{t}\rangle$ denote the mean of the GFPT and the FRT respectively. Results also show that, for the case of $u>1$, while the networks are fractals,  the mean and the variance  of the GFPT scale with the volume of the networks, denoted by $N_t$, as $\langle GFPT_t\rangle \sim N_t^{{2}/{d_s}}$ and $Var(GFPT_t) \sim N_t^{{4}/{d_s}}$, which imply that both the mean and the variance of the GFPT are controlled by the spectral dimension $d_s$. However, for the case of $u=1$, while the networks are nonfractals, the mean and the variance of the GFPT are not controlled by the spectral dimension $d_s=2$, but they are controlled by the transspectral dimension $\tilde{d}_s$ and the mean and the variance of the GFPT scale with the volume of the networks as $\langle GFPT_t\rangle \sim N_t^{{2}/{\tilde{d}_s}}$ and $Var(GFPT_t)\rangle \sim N_t^{{4}/{\tilde{d}_s}}$.

 In order to evaluate the  fluctuation of the  GFPT and FRT, we   calculate  the reduced moments~\cite{HaRo08} of the two random variables and  find that, in the limit of large size (i.e., while $N_t\rightarrow\infty$), the reduced moment of the FRT tends to be infinite, whereas the reduced moment of the GFPT is almost a const.
 Therefore,  on the $(u,v)$ flowers of large size, the FRT  has  huge fluctuation and the estimate provided by its mean is  unreliable, whereas  the fluctuation of the GFPT  is much smaller and the estimate provided by its mean  is more reliable. 

\section{Network model}
\label{sec:TF}

\begin{figure}
\begin{center}
\includegraphics[scale=0.6]{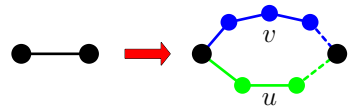}
\caption{Iterative construction method of the $(u,v)$ flowers. The $(u,v)$ flower of generation $t$ $(t>0)$, denoted by $G(t)$,  is obtained from $G(t-1)$ by replacing every edge of $G(t-1)$ by  two parallel paths with lengths $u$  and $v$ ($1\leq u \leq v$) on the right-hand side of the arrow.}
\label{fig:1}       
\end{center}
\end{figure}

\begin{figure}
\begin{center}
\includegraphics[scale=0.6]{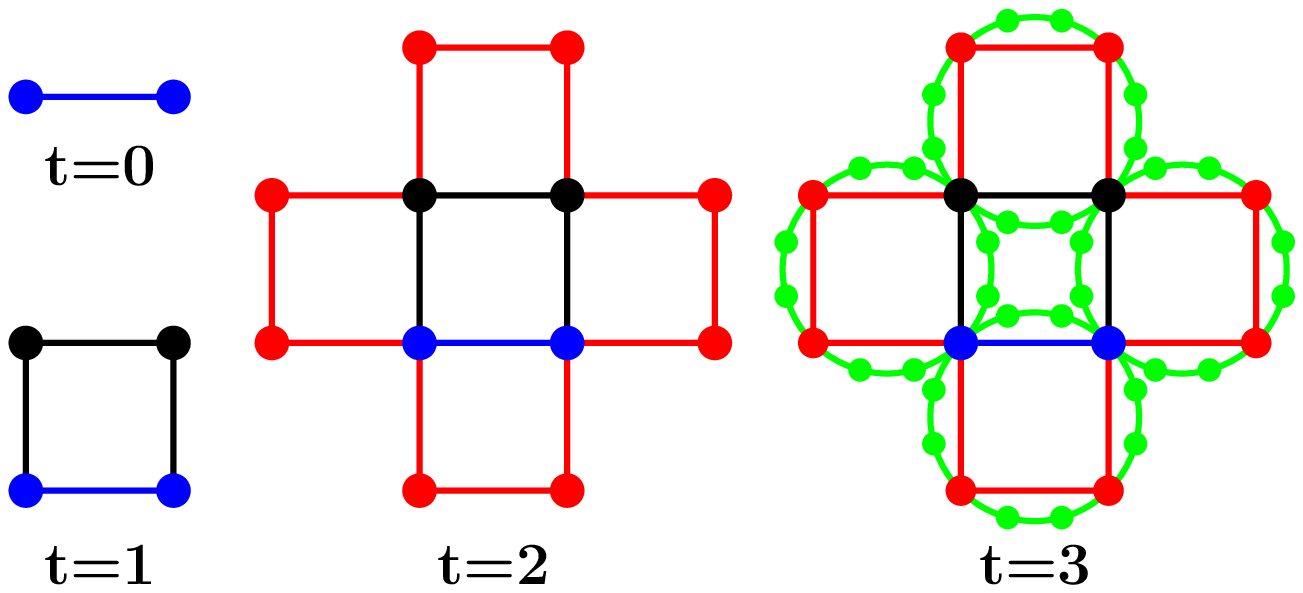}
\caption{The construction of the $(1,3)$ flower with generation $t=0, 1, 2, 3$. }
\label{fig:2}       
\end{center}
\end{figure}

\begin{figure}
\begin{center}
\includegraphics[scale=0.6]{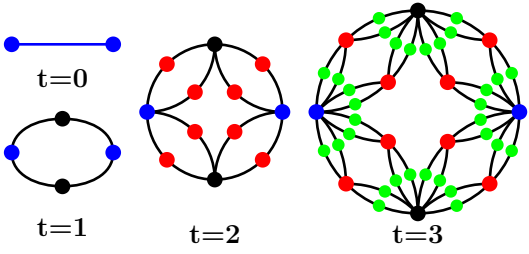}
\caption{The construction of the $(2,2)$ flower with generation $t=0, 1, 2, 3$. }
\label{fig:3}       
\end{center}
\end{figure}

The networks considered here, also called as $(u,v)$ flowers ($1\leq u\leq v$),  are deterministically growing networks which can be  constructed iteratively~\cite{RoHa07}.  Let $G(t)$ denote the $(u,v)$ flower of generation $t$  ($t\geq 0$).  The construction of the $(u,v)$ flower starts from $(t=0)$ two nodes connected by an edge, which corresponds to $G(0)$.  For $t\geq 1$, $G(t)$ is obtained  by replacing every edges of $G(t-1)$   by two parallel paths of $u$ and $v$ edges long.  Fig. \ref{fig:1} shows the iterative construction method of the $(u,v)$ flowers and  Fig. \ref{fig:2},  Fig. \ref{fig:3} show the constructions of the $(1,3)$ flower and the $(2,2)$ flower  with generation $t=0$, $1$, $2$, $3$. Let $w=u+v$. It is easy to see that the total number of edges for $G(t)$ is $E_t = w^{t}$ and the total number of nodes for $G(t)$ is  $N_t=\frac{(w-2)w^{t}+w}{w-1}$~\cite{RoHa07,Zhang11}.

%

The  $(u,v)$ flowers display rich behavior in their topological structure. They follow a power-law degree distribution $P(k)\sim k^{-\gamma}$  with the exponent $\gamma=1+ln(u+v)/ln 2$. For  $u>1$, the networks are \textquotedblleft large-world\textquotedblright and fractals with the fractal dimension $d_f=ln(u+v)/lnu$~\cite{RoHa07}, the walk dimension $d_w=ln(uv)/lnu$ and the spectral dimension $d_s=2ln(u+v)/ln(uv)$~\cite{Hwang10}. For $u=1$, the networks are  small-world and nonfractals with spectral dimension $d_s=2$~\cite{Hwang10}; They have infinite fractal dimension,  walk dimension and finite transfractal dimension $\tilde{d}_f=ln(1+v)/(v-1)$ and transwalk dimension$\tilde{d}_w=ln(v)/(v-1)$. Therefore they are called as transfractals ~\cite{RoHa07}. We can also similarly define their transspectral dimension by $\tilde{d}_s=2\tilde{d}_f/\tilde{d}_w=2ln(1+v)/ln(v)$~\cite{Rammal83}.

\begin{figure}
\begin{center}
\includegraphics[scale=0.6]{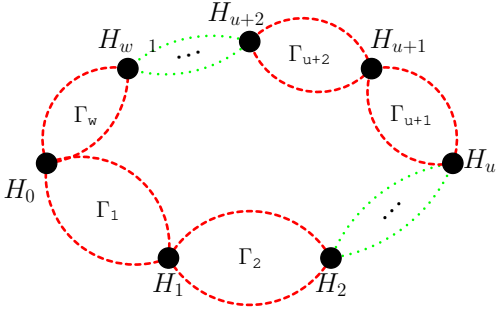}
\caption{Alternative construction of the $(u,v)$ flower which highlights self-similarity: the network of generation $t$, denoted by $G(t)$, is composed of $w\equiv u+v$ copies, called subunits, of $G(t-1)$ which are labeled as  $\Gamma_1$,  $\Gamma_2$,  $\cdots$ ,$\Gamma_w$, and connected to one another at its $w$ hubs, denoted by $H_0$, $H_1$,$\cdots$, $H_{w-1}$.
}
\label{Self_similar}
\end{center}
\end{figure}

The network also has an equivalent  construction method which highlights its self-similarity~\cite{RoHa07}.
Referring to Fig. \ref{Self_similar}, in order to obtain $G(t)$, one can make $w$ copies of $G(t-1)$ and join them  at their hubs (i.e., nodes with the highest degree) denoted by $H_0$, $H_1$,$\cdots$, $H_{w-1}$.  In such a way, $G(t)$ is composed of $w$ copies of $G(t-1)$ labeled as $\Gamma_1$, $\Gamma_2$, $\cdots$, $\Gamma_w$, which are connected with each other by the $w$ hubs. 

\section{Method of exactly calculating the  probability generating function of FRT and GFPT}
\label{PGF_RT_GFPT}

In this section we analyze and present the recurrence relations which the probability generating functions of the FRT for a given  hub and the GFPT to a given hub satisfy. Not loss generality, we  only analyze the FRT for hub $H_0$ and GFPT to hub $H_u$.

Let $F_{i\rightarrow j}(t)$ be  FPT from site $i$ to site $j$ on network $G(t)$. Therefore, $F_{i\rightarrow i}(t)$ is just the first return time on network $G(t)$ for random walk starting from site $i$. Let $P\{ F_{i\rightarrow j}(t)=n\}$ represents the probability that $F_{i\rightarrow j}(t)=n$. Thus, the  probability generating functions of FPT from hub  $H_0$ to hub  $H_u$ is given by
\begin{equation}\label{PGF_FRT}
  \Phi_{FPT}(t,z)=\sum_{n=0}^{+\infty}z^nP\{ F_{H_0\rightarrow H_u}(t)=n\},
\end{equation}
and the  probability generating functions of FRT for hub $H_0$ is given by
\begin{equation}\label{PGF_FRT}
  \Phi_{FRT}(t,z)=\sum_{n=0}^{+\infty}z^nP\{ F_{H_0\rightarrow H_0}(t)=n\}.
\end{equation}

 In the steady state~\cite{LO93}, the probability of finding the random walker at node $i$ is given by $d_i/2E_t$, where $d_i$ is the degree of node $i$. Averaging $F_{i\rightarrow H_u}(t)$ over all starting node $i$,   the GFPT to hub $H_u$ is defined by
\begin{equation}\label{Def_GFPP}
  F_{H_u}(t)=\sum_{i}\frac{d_i}{2E_t}F_{i\rightarrow H_u}(t),
\end{equation}
where the sum runs over all the nodes of $G(t)$.
Therefore the probability generating functions of  GFPT to hub $H_u$ can be expressed as
\begin{equation}\label{PGF_GFPT}
  \Phi_{GFPT}(t,z)=\sum_{n=0}^{+\infty}z^nP\{ F_{H_u}(t)=n\}.
\end{equation}
In order to obtain  $\Phi_{FRT}(t,z)$ and $\Phi_{GFPT}(t,z)$, we first recall their connections with $\Phi_{RT}(t,z)$, which denotes the  probability generating function of the return time for hub $H_0$ in $G(t)$, and is
 defined by
  \begin{equation}\label{PGF_RTO}
 \Phi_{RT}(t,z)=\sum_{n=0}^{+\infty}z^nP(T_{H_0\rightarrow H_0}(t)=n),
\end{equation}
where  $T_{H_0\rightarrow H_0}(t)$ denotes the return time (note: maybe it is not the first time the random walker return to the starting site) on network $G(t)$ for a random walker starting from site $H_0$, and $P(T_{H_0\rightarrow H_0}(t)=n)$ is the  probability that a random walker, starting from site $H_0$, is found at site $H_0$ at time $n$ on network $G(t)$.

It is known that~\cite{HvKa12},  
\begin{equation}\label{R_FRT_RT}
 \Phi_{FRT}(t,z)=1-1/ \Phi_{RT}(t,z)
\end{equation}
and
\begin{equation}\label{R_GFPT_RT}
   \Phi_{GFPT}(t,z)=\frac{z\cdot2^{t-1}}{(1-z)w^{t}}\times\frac{1}{ \Phi_{RT}(t,z)}.
\end{equation}
Therefore, we can calculate  $\Phi_{FRT}(t,z)$ and $\Phi_{GFPT}(t,z)$ if  $\Phi_{RT}(t,z)$ is obtained.

As derived in Appendix~\ref{sec:Rec_relation}, for any $t>0$, $\Phi_{RT}(t,z)$ satisfy the following recurrence relation
\begin{equation}\label{R_RT}
  \Phi_{RT}(t,z)=\Phi_{RT}(t-1,z)/\psi(\Phi_{FPT}(t-1,z)),
\end{equation}
with $\Phi_{RT}(0,z)=\frac{1}{1-z^2}$ and
\begin{equation}\label{Nrat}
\psi(z)\equiv\frac{\Phi_{RT}(0,z)}{\Phi_{RT}(1,z)},
\end{equation}
 where the method to calculate $\Phi_{RT}(1,z)$ is presented in Appendix~\ref{PGF_g1}.

Replacing $\Phi_{RT}(t,z)$ from Eq.~(\ref{R_RT}) in Eqs.~(\ref{R_FRT_RT}) and (\ref{R_GFPT_RT}), for any $t>0$,  we find that $\Phi_{FRT}(t,z)$ and $ \Phi_{GFPT}(t,z)$ satisfy the following recurrence relations,
\begin{equation}\label{R_FRT}
  \Phi_{FRT}(t,z)=1-\psi(\Phi_{FPT}(t-1,z))[1- \Phi_{FRT}(t-1,z)],
\end{equation}
and
\begin{equation}\label{R_GFPT}
  \Phi_{GFPT}(t,z)=\frac{2}{u+v}\psi(\Phi_{FPT}(t-1,z))\Phi_{GFPT}(t-1,z),
\end{equation}
with initial condition $\Phi_{FRT}(0,z)=z^2$, $\Phi_{GFPT}(0,z)=\frac{1}{2}(z+1)z$.

As we have presented  method to calculate $\Phi_{FPT}(t,z)$ for any $t>0$ in Appendix~\ref{PGF_FPTt}. Therefore, we can calculate $\Phi_{FRT}(t,z)$ and $\Phi_{GFPT}(t,z)$  by using Eqs.~(\ref{R_FRT}) and (\ref{R_GFPT}).

\section{Exact calculation of the first and  second moment of FRT}
\label{Sec:M2_FRT}

Calculating the first order derivative with respect to $z$ on both sides of Eq.~(\ref{R_FRT}), we find, for any $t>0$,
\begin{eqnarray}\label{First_d_FRT}
  &&\frac{\partial}{\partial z}\Phi_{FRT}(t,z)\nonumber \\
  &=&\psi(\Phi_{FPT}(t-1,z))\times\frac{\partial}{\partial z}\Phi_{FRT}(t-1,z)\nonumber \\
  &+&\frac{\partial}{\partial x}\left. \psi(x)\right|_{x=\Phi_{FPT}(t-1,z)}\times\frac{\partial}{\partial z}\Phi_{FPT}(t-1,z)\nonumber \\
  &\times& [\Phi_{FRT}(t-1,z)-1].
\end{eqnarray}
Noting that $\Phi_{FRT}(t,1)=1$,  $\Phi_{FPT}(t,1)=1$ and letting  $z=1$ in Eq.~(\ref{First_d_FRT}), we get, for any $t>0$,
\begin{eqnarray}\label{First_m_FRT}
 \langle FRT_t \rangle&=&\left. \frac{\partial}{\partial z}\Phi_{FRT}(t,z)\right|_{z=1}=a_1\times  \langle FRT_{t-1}  \rangle\nonumber \\
&=&\cdots=a_1^t \langle FRT_0  \rangle,
\end{eqnarray}
where $a_1\equiv\psi(1)$ and $ \langle FRT_t \rangle$ denotes the mean first return time of hub $H_0$ on $G(t)$. 
It  is well known that~\cite{LO93}, $\langle FRT_t \rangle$ is controlled by the degree of node $H_0$, and it can also be expressed as
\begin{eqnarray}\label{MFRT}
\langle FRT_t \rangle=\frac{2E_t}{d_{H_0}}=2\times\left(\frac{u+v}{2}\right)^t.
\end{eqnarray}
By comparing Eq.~(\ref{MFRT}) with Eq.~(\ref{First_m_FRT}), we get $a_1=\frac{u+v}{2}$.

Similarity, we can also calculate the second moment of the FRT of hub $H_0$, referred to as $\langle FRT_t^2 \rangle$.
By taking the first order derivative on both sides of Eq.~(\ref{First_d_FRT}), we obtain
   \begin{eqnarray}\label{Second_d_FRT}
     &&\frac{\partial^2}{\partial z^2}\Phi_{FRT}(t,z)\nonumber \\
     &=&\psi(\Phi_{FPT}(t-1,z))\times\frac{\partial^2}{\partial z^2}\Phi_{FRT}(t-1,z)\nonumber \\
  &+&2\frac{\partial}{\partial x}\left.\psi(x)\right|_{x=\Phi_{FPT}(t-1,z)}\times\frac{\partial}{\partial z}\Phi_{FPT}(t-1,z)\nonumber \\
  &\times&\frac{\partial}{\partial z}\Phi_{FRT}(t-1,z)+{[\Phi_{FRT}(t-1,z)-1]}\nonumber \\
  &\times&\frac{\partial^2}{\partial z^2}\left.\psi(x)\right|_{x=\Phi_{FPT}(t-1,z)}\times\frac{\partial}{\partial z}\Phi_{FPT}(t-1,z).
  \end{eqnarray}
By posing $z=1$ in  Eq.~(\ref{Second_d_FRT}), we get
 \begin{eqnarray}\label{D2_FRT}
      &&\left. \frac{\partial^2}{\partial z^2}\Phi_{FRT}(t,z)\right|_{z=1}\nonumber\\
      &=&\left.a_1\times\frac{\partial^2}{\partial z^2}\Phi_{FRT}(t-1,z)\right|_{z=1}\nonumber\\
      &+&2a_2\times\langle FRT_{t-1}  \rangle\times\langle FPT_{t-1}  \rangle,
  \end{eqnarray}
  where $a_1\equiv\psi(1)$, $a_2\equiv\frac{\partial}{\partial z}\left.\psi(z)\right|_{z=1}$ and $ \langle FPT_t \rangle$ denotes the mean first passage time from hub $H_0$ to hub $H_u$ on $G(t)$. Therefore, the second moment of FRT of node $H_0$  satisfies
  \begin{eqnarray}\label{R_Second_m_FRT}
      & &\langle FRT^2_t\rangle=
      \left. \frac{\partial^2}{\partial z^2}\Phi_{FRT}(t,z)\right|_{z=1}+\langle FRT_t \rangle \nonumber\\
      &=&a_1\langle FRT^2_{t-1}  \rangle +2a_2\langle FRT_{t-1}  \rangle\langle FPT_{t-1}  \rangle.
  \end{eqnarray}
Using Eq.~(\ref{R_Second_m_FRT}) recursively, and inserting Eqs.~(\ref{First_m_FRT}) and (\ref {MFPT}) into it, we get
  \begin{eqnarray}\label{Second_m_FRT}
     \langle FRT^2_t\rangle
     &=&a_1^t\langle FRT^2_{0}  \rangle +2a_2\langle FRT_{t-1}  \rangle \langle FPT_{t-1}\rangle\nonumber \\
     &\times&\left[ 1+\frac{1}{\langle FPT_1  \rangle}+\cdots+\frac{1}{\langle FPT_1  \rangle^{t-1}} \right] \nonumber\\
     &=&a_1^t\langle FRT^2_{0}  \rangle +2\frac{a_2}{a_1\langle FPT_1  \rangle}\langle FRT_{t}  \rangle \langle FPT_{t}\rangle\nonumber \\
     &\times&\frac{1-\frac{1}{\langle FPT_1  \rangle^t}}{1-\frac{1}{\langle FPT_1  \rangle}},
  \end{eqnarray}
  which shows that
  \begin{equation}\label{S_M2FRT}
   \langle FRT^2_t\rangle \sim \langle FRT_{t}  \rangle \langle FPT_{t}\rangle.
\end{equation}
Therefore the variance of the FRT, denoted by $Var(FRT_t)$, satisfies
  \begin{eqnarray}\label{Var_FRT}
   Var(FRT_t)&=&\langle FRT^2_t\rangle-\langle FRT_t\rangle^2\nonumber\\
     &\sim& \langle FRT_{t}  \rangle \langle FPT_{t}\rangle\times[1-\frac{FRT_{t}}{FPT_{t}}].
\end{eqnarray}
As shown in Eqs.~(\ref{MFPT_E}), (\ref{MFRT}), and (\ref{GMFPT}), $\langle FPT_t\rangle= (uv)^t$, $\langle FRT_t\rangle=2(\frac{u+v}{2})^t$ and $\langle GFPT_t\rangle \sim (uv)^t$. Therefore $\langle GFPT_t\rangle \sim\langle FPT_t\rangle$ and $\frac{FRT_{t}}{FPT_{t}}\longrightarrow 0$ while $t \longrightarrow \infty$. Thus, for networks of large size,
  \begin{eqnarray}\label{S_Var_FRT}
   Var(FRT_t)\sim\langle FRT_{t}  \rangle \langle GFPT_{t}\rangle.
\end{eqnarray}
As discussed in Sec.\ref{Sec:M2_GFPT},  $\langle GFPT_{t}\rangle$ is controlled by  the spectral (or transspectral) dimension of the network. Therefore Eqs.~(\ref{S_Var_FRT}) shows that $Var(FRT_t)$ is controlled by  the degree of node $H_0$ and the spectral (or transspectral) dimension of the network.

  By determining every parameters in Eq.~(\ref{Second_m_FRT}), we can further obtain the exactly formula of $\langle FRT^2_t\rangle$ and $Var(FRT_t)$.
   Recalling that $\Phi_{FRT}(0,z)=z^2$ and calculating the first and second order derivative of $\Phi_{FRT}(0,z)$ with respect to $z$ and fixing $z=1$, we obtain $\langle FRT_0\rangle=2$ and
  \begin{eqnarray}\label{Second_m_g0}
      \langle FRT^2_0\rangle=4.
  \end{eqnarray}
 For the parameter $a_2$, it can be obtained by calculating the first order derivative of $\psi(z)$ with respect to $z$ and fixing $z=1$. 
Therefore  $\langle FRT^2_t\rangle$ can be exactly determined by using Eq.~(\ref{Second_m_FRT}) and $Var(FRT_t)$ can also be calculated by $Var(FRT_t)=\langle FRT^2_t\rangle-\langle FRT_t\rangle^2$.

Here we take  the specific case of $u=2$ and $v=2$ as an example. As derived in Appendix~\ref{PGF_g1}, $\psi(z)=\frac{2}{2-z^2}$.
Thus $a_2=\frac{\partial}{\partial z}\left.\psi(z)\right|_{z=1}=4$ and all parameters in Eq.~(\ref{Second_m_FRT}) are known. 
Therefore $\langle FRT^2_t\rangle= \frac{8^{t+1}}{3}+\frac{2^{t+2}}{3}$ and
\begin{eqnarray}\label{2_2_Var_FRT}
     Var(FRT_t)&=& \frac{8^{t+1}}{3}+\frac{2^{t+2}}{3}-4^{t+1}.
  \end{eqnarray}

In order to evaluate the  fluctuation of the FRT, we  calculate  the reduced moment~\cite{HaRo08}, defined by $R(FRT_t)=\sqrt{Var(FRT_t)} / \langle FRT_t\rangle$. We  find it grows with the increasing of the network and in the limit of large size (i.e., while $t\rightarrow\infty$),
 \begin{eqnarray}\label{Reduce_m_FRT}
    R(FRT_t)
            &\sim &\left(\frac{2uv}{u+v}\right)^{t/2}\rightarrow\infty,
\end{eqnarray}
 which shows that, on the $(u,v)$ flower of large size, the FRT of hub node  has  huge fluctuation and the estimate provided by MFRT is unreliable.%

\section{Exact calculation of the first and second moment of the GFPT}
\label{Sec:M2_GFPT}

By calculating the first order derivative with respect to $z$ on both sides of Eq.~(\ref{R_GFPT}), for any $t>0$, we find,
\begin{eqnarray}\label{First_d_GFPT}
  &&\frac{\partial}{\partial z}\Phi_{GFPT}(t,z)\nonumber \\
  &=&\frac{2}{u+v}\left\{\psi(\Phi_{FPT}(t-1,z))\times\frac{\partial}{\partial z}\Phi_{GFPT}(t-1,z)\right.\nonumber \\
  &+&\frac{\partial}{\partial x}\left. \psi(x)\right|_{x=\Phi_{FPT}(t-1,z)}\times\frac{\partial}{\partial z}\Phi_{FPT}(t-1,z)\nonumber \\
  &\times&\left. \Phi_{GFPT}(t-1,z)\right\}.
\end{eqnarray}
Noting that $\Phi_{GFPT}(t,1)=1$,  $\Phi_{FPT}(t,1)=1$, $\psi(1)=\frac{u+v}{2}$ and posing  $z=1$ in Eq.~(\ref{First_d_GFPT}), we get, for any $t>0$,
\begin{eqnarray}\label{R_First_m_GFPT}
 \langle GFPT_t \rangle&=&\left. \frac{\partial}{\partial z}\Phi_{GFPT}(t,z)\right|_{z=1}\nonumber \\
&=& \langle GFPT_{t-1}\rangle+\frac{2 a_2}{u+v} \langle FPT_{t-1}\rangle,
\end{eqnarray}
where $a_2$ is the same as that of Eq.~(\ref{D2_FRT}) and $ \langle GFPT_t \rangle$ denotes the mean of global first passage time to hub $H_u$ on $G(t)$. 

Using Eq.~(\ref{R_First_m_GFPT}) recursively, and replacing $\langle FPT_{k}\rangle$ from Eq.~(\ref {MFPT}),
\begin{eqnarray}\label{First_m_GFPT}
 \langle GFPT_t \rangle&=&\langle GFPT_{t-1}\rangle+\frac{2 a_2}{u+v} \langle FPT_{t-1}\rangle \nonumber \\
                   &=&\langle GFPT_{t-2}\rangle+\frac{2 a_2}{u+v} \langle FPT_{t-2}\rangle+\frac{2 a_2}{u+v} \langle FPT_{t-1}\rangle \nonumber \\
                   &=&\cdots\nonumber \\
                   &=&\langle GFPT_{0}\rangle+\frac{2 a_2}{u+v}\left[ \langle FPT_{0}\rangle+\langle FPT_{1}\rangle+\cdots\right.\nonumber \\
                   &+&\left. \langle FPT_{t-1}\rangle \right]\nonumber \\
                   &=&\langle GFPT_{0}\rangle+\frac{2 a_2}{u+v}\left[ 1+uv+\cdots+(uv)^{t-1}\right]\nonumber \\
                   &=&\frac{3}{2}+\frac{2 a_2}{u+v}(uv)^{t-1}\frac{1-(uv)^{-t}}{1-(uv)^{-1}},
\end{eqnarray}
where $\langle GFPT_{0}\rangle=\frac{3}{2}$, which can be easily calculated on the network of generation $0$.

Similarity, we can also calculate the second moment of the GFPT to hub $H_u$, referred to as $\langle GFPT_t^2 \rangle$.
By taking the first order derivative on both sides of Eq.~(\ref{First_d_GFPT}), we obtain, for any $t>0$,
   \begin{eqnarray}\label{Second_d_GFPT}
     &&\frac{\partial^2}{\partial z^2}\Phi_{GFPT}(t,z)\nonumber \\
     &=&\frac{2}{u+v}\left\{\psi(\Phi_{FPT}(t-1,z))\times\frac{\partial^2}{\partial z^2}\Phi_{GFPT}(t-1,z)\right.\nonumber \\
  &+&2\frac{\partial}{\partial x}\left.\psi(x)\right|_{x=\Phi_{FPT}(t-1,z)}\times\frac{\partial}{\partial z}\Phi_{FPT}(t-1,z)\nonumber \\
  &\times&\frac{\partial}{\partial z}\Phi_{GFPT}(t-1,z)+\Phi_{GFPT}(t-1,z)\nonumber \\
  &\times&\left[\frac{\partial^2}{\partial x^2}\left.\psi(x)\right|_{x=\Phi_{FPT}(t-1,z)}\times\left(\frac{\partial}{\partial z}\Phi_{FPT}(t-1,z)\right)^2\right.\nonumber \\
  &+&\left.\left.\frac{\partial}{\partial x}\left.\psi(x)\right|_{x=\Phi_{FPT}(t-1,z)}\times\frac{\partial^2}{\partial z^2}\Phi_{FPT}(t-1,z)\right]\right\}.
  \end{eqnarray}
By posing $z=1$ in  Eq.~(\ref{Second_d_GFPT}), for any $t>0$, we get
 \begin{eqnarray}\label{D2_GFPT}
      &&\left. \frac{\partial^2}{\partial z^2}\Phi_{GFPT}(t,z)\right|_{z=1}\nonumber\\
      &=&\frac{2}{u+v}\left\{\left.a_1\times\frac{\partial^2}{\partial z^2}\Phi_{GFPT}(t-1,z)\right|_{z=1}\right.\nonumber\\
      &+&2a_2\langle GFPT_{t-1}  \rangle\langle FPT_{t-1}  \rangle +a_3(\langle FPT_{t-1}  \rangle)^2\nonumber\\
      &+&\left.a_2\times\left.\frac{\partial^2}{\partial z^2}\Phi_{FPT}(t-1,z)\right|_{z=1}\right\},
  \end{eqnarray}
  where $a_1=\frac{u+v}{2}$, $a_2$ are the same as that of Eq.~(\ref{D2_FRT}) and $a_3\equiv\frac{\partial^2}{\partial z^2}\left.\psi(z)\right|_{z=1}$. Therefore, for any $t>0$,
  \begin{eqnarray}\label{R_Second_m_GFPT}
      & &\langle GFPT^2_t\rangle=
      \left. \frac{\partial^2}{\partial z^2}\Phi_{GFPT}(t,z)\right|_{z=1}+\langle GFPT_t \rangle \nonumber\\
      &=&\langle GFPT^2_{t-1}  \rangle +\frac{4a_2}{u+v}\langle GFPT_{t-1}  \rangle\langle FPT_{t-1}  \rangle\nonumber\\
      &+&\frac{2a_3}{u+v}(\langle FPT_{t-1}  \rangle)^2+\frac{2a_2}{u+v}\langle FPT^2_{t-1}  \rangle,
  \end{eqnarray}
  where $\langle FPT^2_{t}  \rangle$ denotes the second moment of the  first passage time from hub $H_0$ to hub $H_u$ on $G(t)$.

 Replacing $\langle GFPT_{t-1} \rangle$, $\langle FPT_{t-1}  \rangle$ and $\langle FPT^2_{t-1}  \rangle$ from Eqs.~(\ref{First_m_GFPT}), (\ref{MFPT}) and (\ref{M2FPT}) respectively in Eq.~(\ref{R_Second_m_GFPT}), for any $t>0$,
   \begin{eqnarray}\label{R_M2_GFPT}
      \langle GFPT^2_t\rangle
      &=&\langle GFPT^2_{t-1}  \rangle +\frac{4a_2}{u+v}\langle GFPT_{t-1}  \rangle\langle FPT_{t-1}  \rangle\nonumber\\
      &+&\frac{2a_3}{u+v}(\langle FPT_{t-1}  \rangle)^2+\frac{2a_2}{u+v}\langle FPT^2_{t-1}  \rangle \nonumber\\
      &=&\langle GFPT^2_{t-1}  \rangle +k_1(uv)^{2t-2}+k_2 (uv)^{t-1},
  \end{eqnarray}
  where
 \begin{equation} \label{K_1}
   k_1= \frac{8a_2^2}{w^2(uv-1)}+\frac{2a_3}{u+v}+\frac{2a_2(\langle FPT_{1}^2  \rangle-\langle FPT_{1}  \rangle)}{wuv(uv-1)},
   \end{equation}
    and
   \begin{equation} \label{K_2}
   k_2= \frac{8a_2}{u+v}-\frac{8a_2^2}{w^2(uv-1)}-\frac{2a_2(\langle FPT_{1}^2  \rangle-\langle FPT_{1}  \rangle)}{wuv(uv-1)}.
   \end{equation}

Using Eq.~(\ref{R_M2_GFPT}) recursively, for any $t>0$,
  \begin{eqnarray}\label{Second_m_GFPT}
      \langle GFPT^2_t\rangle
      &=&\langle GFPT^2_{t-1}  \rangle +k_1(uv)^{2t-2}+k_2 (uv)^{t-1}\nonumber\\
      &=&\langle GFPT^2_{t-2}  \rangle +k_1(uv)^{2t-4}+k_2 (uv)^{t-2}\nonumber\\
      &&+k_1(uv)^{2t-2}+k_2 (uv)^{t-1}\nonumber\\
      &=&\cdots\nonumber\\
      &=&\langle GFPT^2_{0}  \rangle +k_1\left[1+(uv)^{2}+\cdots+ (uv)^{2t-2}\right]\nonumber\\
      &&+k_2\left[1+uv+\cdots+ (uv)^{t-1}\right]\nonumber\\
      &=&\langle GFPT^2_{0} \rangle\! +\!k_1\frac{(uv)^{2t}\!-\!1}{(uv)^2-1}\! +\!k_2\frac{(uv)^t\! -\!1}{uv-1}.
  \end{eqnarray}
Because the network of generation $0$ is just an edge, it is easy to obtain $\langle GFPT^2_0\rangle=\frac{5}{2}$.  Given $u$ and $v$, we can derive $\Phi_{RT}(1,z)$ and $\Phi_{FPT}(1,z)$ by using the method presented in Appendix~\ref{PGF_g1}. Then $\psi(z)$,   $a_2$, $a_3$, $\langle FPT_{1}  \rangle$ and $\langle FPT_{1}^2  \rangle$ can  all be obtained. Thus $k_1$ and $k_2$ is obtained.  Therefore $\langle GFPT^2_t\rangle$ can be exactly calculated and the the variance of the GFPT,  denoted by $Var(GFPT_t)$, can also be calculated exactly by $Var(GFPT_t)=\langle GFPT^2_t\rangle-[\langle GFPT_t\rangle]^2$.

Here we take  the specific case of $u=2$ and $v=2$ as an example. As derived in Appendix~\ref{PGF_g1}, $\psi(z)=\frac{2}{2-z^2}$ and $\Phi_{FPT}(1,z)=\frac{-z^2}{z^2-2}$.
Thus $a_2=\frac{\partial}{\partial z}\left.\psi(z)\right|_{z=1}=4$, $a_3=\frac{\partial^2}{\partial z^2}\left.\psi(z)\right|_{z=1}=20$, and
$\langle FPT_{1}^2  \rangle-\langle FPT_{1}  \rangle=\frac{\partial^2}{\partial z^2}\left.\Phi_{FPT}(1,z)\right|_{z=1}=20$. Inserting all parameters into Eqs.~(\ref{K_1}), (\ref{K_2}) (\ref{First_m_GFPT}) and (\ref{Second_m_GFPT}), we obtain  $k_1=16$, $k_2=2$, $\langle GFPT_t\rangle= \frac{2}{3}\times4^{t}+\frac{5}{6}$,  $\langle GFPT^2_t\rangle= \frac{16^{t+1}}{15}+\frac{2}{3}\times4^{t}+\frac{23}{30}$ and
\begin{eqnarray}\label{2_2_Var_GFPT}
     Var(GFPT_t)= \frac{28}{45}\times16^{t}-\frac{1}{9}\times4^{t+1}+\frac{13}{180}.
  \end{eqnarray}

We can also find from Eqs.~(\ref{First_m_GFPT}) and (\ref{Second_m_GFPT})  that $\langle GFPT_t\rangle \sim (uv)^t$ and $\langle GFPT_t\rangle \sim (uv)^{2t}$. Therefore 
\begin{equation}\label{Var_GFPT}
   Var(GFPT_t)\sim  \langle GFPT_{t}\rangle^2 \sim  (uv)^{2t}.
\end{equation}

Since the volume of the $(u,v)$ flower scales as $N_t \sim (u+v)^t$ for large sizes,  
\begin{equation}\label{GMFPT}
   \langle GFPT_t\rangle\sim N_t^{\frac{ln(uv)}{ln(u+v)}},
\end{equation}
and  
\begin{equation}\label{S_Var_GFPT}
  Var(GFPT_t) \sim N_t^{\frac{2ln(uv)}{ln(u+v)}}.
\end{equation}

For the case of $u>1$, the $(u,v)$ flowers have spectral dimension $d_s=\frac{2ln(u+v)}{ln(uv)}$~\cite{Hwang10}.  Therefore Eqs.~(\ref{GMFPT}) and (\ref{S_Var_GFPT}) show that,  $\langle GFPT_t\rangle \sim N_t^{{2}/{d_s}}$ and $Var(GFPT_t) \sim N_t^{{4}/{d_s}}$. Both the mean and variance of the GFPT are controlled by the spectral dimension $d_s$.

For the case of $u=1$, the $(u,v)$ flowers are nonfractal, they have  spectral dimension $d_s=2$~\cite{Hwang10}. Our results show that both the mean and variance of the GFPT have no direct relation with $d_s$. Although the $(1,v)$ flowers are not fractals , they are transfractals with the transfractal dimension $\tilde{d}_f=ln(1+v)/(v-1)$ and the transwalk dimension $\tilde{d}_w=ln(v)/(v-1)$~\cite{RoHa07}. Similar to  ${d}_s=2{d}_f/{d}_w$,  we  define the transspectral dimension by $\tilde{d}_s=2\tilde{d}_f/\tilde{d}_w=2ln(1+v)/ln(v)$~\cite{Rammal83}, Eqs.~(\ref{GMFPT}) and (\ref{S_Var_GFPT}) show that,  $\langle GFPT_t\rangle \sim N_t^{{2}/{\tilde{d}_s}}$ and $Var(GFPT_t) \sim N_t^{{4}/{\tilde{d}_s}}$. Therefore the mean and variance of the FRT are controlled by the transspectral dimension $\tilde{d}_s$.

In order to evaluate the  fluctuation of the GFPT, we can further calculate  the reduced moment of GFPT~\cite{HaRo08}, defined by $R(GFPT_t)=\sqrt{Var(GFPT_t)} / \langle GFPT_t\rangle$. Result shows that it grows with the increasing of the network and in the limit of large size (i.e., while $t\rightarrow\infty$),
 \begin{eqnarray}\label{Reduce_m_GFPT}
    R(GFPT_t)
            &\approx&\sqrt{\frac{k_1w^2(uv-1)^2}{4a_2^2(u^2v^2-1)}-1}.
\end{eqnarray}
 By comparing the result with that of the FRT, the fluctuation of the GFPT is much smaller  and the estimate provided by its mean is more reliable.

\section{Conclusions}
\label{sec:4}

 We have presented  method to calculate exactly the mean and the variance of the GFPT and FRT for a given hub  on the $(u,v)$ flowers and the scaling behavior of  mean and the variance of the GFPT and FRT  were also analyzed.  Results show that, for the case of $u>1$, while the networks are fractals,  both the mean and the variance of the GFPT are controlled by the spectral dimension $d_s$. However, for the case of $u=1$, while the networks are nonfractals, the mean and the variance of the GFPT are not controlled by the spectral dimension, but they are controlled by the transspectral dimension.   Results also show that  the variance  of the GFPT and FRT scale with the mean  of the GFPT and FRT as $Var(GFPT_t) \sim  \langle GFPT_{t}\rangle^2$ and $Var(FRT_t) \sim \langle FRT_{t}  \rangle \langle GFPT_{t}\rangle$. Note that the mean of the FRT  is controlled by the degree of the node. We found that the variance of the  FRT  are controlled by the degree of the node and the spectral dimension (or transspectral dimension) of the network.

 We have also  calculated  the reduced moments of the GFPT and FRT and  find that, in the limit of large size, the reduced moment of the FRT tends to be infinite, whereas the reduced moment of the GFPT is almost a const.
 Therefore,  on the $(u,v)$ flowers of large size, the FRT  has  huge fluctuation and the estimate provided by its mean is  unreliable, whereas  the fluctuation of the GFPT  is much smaller and the estimate provided by its mean  is more reliable.

Of course, the method proposed here  can also be used on other self-similar graph such as Sierpinski gaskets, tree-like fractals, recursive  scale-free trees and etc.

\begin{acknowledgments}
This work was supported  by the  research project of the national science and technology thought library of Guangzhou  under Grant No. 2016sx010.
\end{acknowledgments}

\appendix

  \section{Probability generating function and its properties}
\label{sec:PGF}
  Let $T$ be a discrete random variable which takes only non-negative integer values, and whose probability distribution is $p_k$ ($k=0,1,2,\cdots$) . The probability generating function of $T$ is defined as 
 \begin{equation}\label{Def_PGF}
   \Phi_{T}(z)=\sum_{k=0}^{+\infty}z^k p_k.
 \end{equation}
 The probability generating function of $T$ is  determined by the probability distribution and, in turn, it uniquely determines the probability distribution. If $T_1$ and $T_2$ are two random variables with the same probability generating function, then they have the same probability distribution. Given the probability generating function $\Phi_{T}(z)$ of the random variable $T$, we can obtain the probability distribution $p_k$ ($k=0,1,2,\cdots$)  as the coefficient of $z^k$ in the Taylor's series expansion of $\Phi_{T}(z)$ about $z=0$. 

Also, the $n$-th moment $\langle T^n \rangle \equiv \sum_{k=0}^{+\infty} k^n p_k$, can be written in terms of combinations of derivatives (up to the $n$-th order) of $\Phi_{T}(z)$ calculated in $z=1$. In particular,
\begin{eqnarray}\label{n_moment}
\langle T \rangle  &=& \frac{\partial \Phi_{T}(z)}{ \partial z} \Big |_{z=1},\\
\langle T^2 \rangle  &=&  \frac{\partial^2 \Phi_{T}(z)}{ \partial z^2} \Big|_{z=1} + \frac{\partial \Phi_{T}(z)}{ \partial z} \Big|_{z=1}.
 \end{eqnarray}

Finally, we list  some useful properties of the probability generating function~\cite{Gut05}, which would be useful in this paper:
   \begin{itemize}
  \item Let $T_1$ and $T_2$ be two independent random variables with probability generating functions $\Phi_{T_1}(z)$ and $\Phi_{T_2}(z)$, respectively. Then, the probability generating function of random variable $T_1+T_2$ reads as
  \begin{equation}\label{Sum_PGF2}
    \Phi_{T_1+T_2}(z)=\Phi_{T_1}(z)\Phi_{T_2}(z).
 \end{equation}
  \item Let $N$, $T_1$, $T_2$, $\cdots$ be independent random variables. If $T_i$ ($i=1, 2, \cdots$) are identically distributed, each with probability generating function $\Phi_{T}(z)$, and, being $\Phi_{N}(z)$ the probability generating function of $N$, the random variable defined as
 \begin{equation}\label{Sum_R_V}
    S_N=T_1+T_2+\cdots+T_N
 \end{equation}
 has probability generating function
  \begin{equation}\label{Sum_PGFn}
    \Phi_{S_N}(z)=\Phi_{N}(\Phi_{T}(z)).
 \end{equation}
 \end{itemize}

\section{Probability generating function of FPT and return time on the networks of generation $1$  }
\label{PGF_g1}
The structure of $(u$, $v)$ flower of generation $1$ is a ring with $w=u+v$ nodes, which are labeled as $H_0$, $H_1$, $H_2$, $\cdots$, $H_{w-1}$. In this appendix, we present method to calculate  $\Phi_{RT}(1,z)$ (i.e., the probability generating function (PGF) of the return time for hub $H_0$), $\Phi_{FPT}(1,z)$ (i.e., the PGF of the FPT from $H_0$ to $H_u$) and $\Phi^a_{RT}(1,z)$ (i.e., the PGF of the return time for hub $H_0$ in the presence of a trap $H_u$).

Let
 $$M=(P_{ij})_{w\times w}$$
  be the  transition probability matrix for random walks on the (u,v) flower of generation $1$. Here
\begin{equation}
\label{Pij}
  P_{ij}=\left\{ \begin{array}{ll} \frac{1}{d_i} & \text{if $H_i\sim H_j$, and $x$ is not a trap}\\ 0 & \textrm{others} \end{array} \right.,
\end{equation}
where $H_i\sim H_j$ means that there is an edge between $H_i$ and $H_j$ and $d_i$ is the degree of node $H_i$.
Then we can calculate the probability generating function  directly by
 \begin{eqnarray}\label{Formular_PGF}
\Psi(z)=\sum_{n=0}^{+\infty}(zM)^n=(I-zM)^{-1},
\end{eqnarray}
where $\Psi(z)=(\Psi_{ij}(z))_{w \times w}$ and  $\Psi_{ij}(z)$ is the PGF of passage time from node $H_i$ to $H_j$, whereas $\Psi_{ii}(z)$ is the PGF of return time of node  $H_i$.  If $H_j$ is a trap, $\Psi_{ij}(z)$ is just the PGF of first passage time from node $H_i$ to $H_j$ and $\Psi_{ii}(z)$ is just the PGF of the return time in the presence of a trap $H_j$ for random walkers starting from node $H_i$.

\paragraph*{Exact calculation of  $\Phi_{RT}(1,z)$ and $\psi(z)$ on $(2$, $2)$ flower and $(1$, $3)$ flower.}

For network of generation $1$, the structure of $(2$, $2)$ flower and $(1$, $3)$ flower are the same. Therefore $\Phi_{RT}(1,z)$ is the equal for $(2$, $2)$ flower and $(1$, $3)$ flower.
In order to calculate $\Phi_{RT}(1,z)$,   no trap  is introduced in the network. Therefore,
\begin{equation}
\label{PM1}
  M=\left( \begin{array}{llll}
  0   &  \frac{1}{2}    & 0   & \frac{1}{2}\\
  \frac{1}{2} & 0  & \frac{1}{2} & 0 \\
  0  & \frac{1}{2}  & 0  & \frac{1}{2}  \\
  \frac{1}{2} & 0  & \frac{1}{2} & 0
   \end{array} \right).
\end{equation}
Replacing $M$ from Eq.~(\ref{PM1}) in Eq.~(\ref{Formular_PGF}), calculating $(I-zM)^{-1}$ by using the tools of \emph{Matlab}, we obtain $\Psi(z)$ for this case.  Then the probability generating function of the RT of hub $H_0$  is
\begin{equation}\label{PhiRT0}
 \Phi_{RT}(1,z)=\Psi_{00}(z)=\frac{z^2-2}{2(z^2-1)}.
\end{equation}
Notice that $\Phi_{RT}(0,z)=\frac{1}{1-z^2}$, we obtain
\begin{equation}\label{2_2_Nrat}
   \psi(z)\equiv\frac{\Phi_{RT}(0,z)}{\Phi_{RT}(1,z)}=\frac{2}{2-z^2}.
\end{equation}

\paragraph*{Exact calculation of  $\Phi_{FPT}(1,z)$ and $\Phi^a_{RT}(1,z)$ on  $(2$, $2)$ flower and $(1$, $3)$ flower.}
Firstly we calculate $\Phi_{FPPT}(1,z)$, $\Phi^a_{RT}(1,z)$ on  $(2$, $2)$ flower.  Let hub $H_2$ be a trap. Therefore,
\begin{equation}
\label{PM2}
  M=\left( \begin{array}{llll}
0   &  \frac{1}{2}    & 0   & \frac{1}{2}\\
  \frac{1}{2} & 0  & \frac{1}{2} & 0 \\
  0  & 0  & 0  & 0  \\
  \frac{1}{2} & 0  & \frac{1}{2} & 0
   \end{array} \right).
\end{equation}
 Calculating $\Psi(z)$ from Eq.~(\ref{Formular_PGF}) by using \emph{Matlab},  the probability generating function of the FPT from $H_0$ to $H_u$ on  $(2$, $2)$ flower is
\begin{equation}\label{PhiFPT0}
 \Phi_{FPT}(1,z)= \Psi_{02}(z)=\frac{-z^2}{z^2-2},
\end{equation}
and the probability generating function of the return time of hub $H_0$ in the presence of a trap $H_2$ on $(2$, $2)$ flower reads as
\begin{equation}\label{PhiaRT0}
  \Phi_{RT}^a(1,z)= \Psi_{00}(z)=\frac{-2}{z^2-2}.
\end{equation}

Furthermore, we can also derive $\Phi_{FPPT}(1,z)$, $\Phi^a_{RT}(1,z)$ for  $(1$, $3)$ flower from $\Psi(z)$ we just get. The results are
\begin{equation}\label{PhiFPT0}
 \Phi_{FPT}(1,z)= \Psi_{12}(z)=\frac{-z}{z^2-2},
\end{equation}
and
\begin{equation}\label{PhiaRT0}
  \Phi_{RT}^a(1,z)= \Psi_{11}(z)=\frac{z^2-4}{2(z^2-2)}.
\end{equation}
Note: the structure of $(1$, $3)$ flower of generation $1$ is a ring with $4$ nodes, which is the same as that of $(2$, $2)$ flower. The first passage time from $H_0$ to $H_3$ is the same as the first passage time from $H_1$ to $H_2$ and the return time of hub $H_0$ in the presence of a trap $H_3$ is the same as the return time of hub $H_1$ in the presence of a trap $H_2$.

\section{Derivation of Eq.~(\ref{R_RT})}
 \label{sec:Rec_relation}

Considering any return path $\pi$  starting from $H_0$ and ending at $H_0$ on $G(t)$. Its length is just the return time  $T_{H_0\rightarrow H_0}(t)$.
  Let  $v_i$ be the node of $G(t)$ reached at time $i$. Then the path can be rewritten as $$\pi=(v_0=H_0,v_1,v_2\cdots,v_{T_{H_0\rightarrow H_0}(t)}=H_0).$$
In general,  $H_0$ and $H_u$ would appear in the path $\pi$ for many times.
We denote with $\Omega$ the set of nodes $\{H_0,H_u \}$ and introduce the observable $\tau_i=\tau_i(\pi)$,
which is defined recursively as follows:
\begin{eqnarray}
\tau_0(\pi) &=& 0,\label{OB1}\\
\tau_i(\pi) &=&\min\{k: k>\tau_{i-1},v_{\tau_i}\in \Omega, v_{\tau_i}\neq v_{\tau_{i-1}} \}  \label{OB2}.
\end{eqnarray}
Then,  considering only nodes in the set $\Omega$, the path $\pi$ can be restated into a ``simplified path'' defined as
\begin{equation}
\label{Def_simp}
  \iota(\pi)=(v_{\tau_{0}}=H_0,v_{\tau_1},\cdots, v_{\tau_L}=H_0 ),
\end{equation}
where  $L=\max\{i: v_{\tau_i}=H_0\}$, which is the total number of observables obtained from the path $\pi$. In fact, the ``simplified path'' is obtained by removing any nodes of $\pi$ except $H_0$ and $H_u$. If node $H_0$ (or $H_u$) appears consecutively in the ``simplified path'', we just record the first one.

Note that $\Omega$ represents  the nodes set of the (u,v) flower with generation $0$ and the path $\iota(\pi)$ includes only the  nodes of $\Omega$. Thus, $\iota(\pi)$ is just a return path of $H_0$  on the  (u,v) flower with generation $0$ and $L$ is just the return time of of $H_0$  on the  (u,v) flower with generation $0$. Therefore, the probability generating function of $L$ is $ \Phi_{RT}(0,z)$.

For any return path  $\pi$ of $H_0$,   maybe $v_{\tau_L}$ is not the last node of path  $\pi$. That is to say, after node $v_{\tau_L}$, path  $\pi$ includes a sub-path from $H_0$ to $H_0$, which does not reach  $H_u$. In principle, the sub-path may include any node of $G(t)$ except $H_u$. Therefore, the sub-path can be regarded as a return path of $H_0$ in the presence of an absorbing hub $H_u$.  We denote its length by $T^a_{H_0\rightarrow H_0}(t)$  and denote its probability generating function by $ \Phi^a_{RT}(t,z)$.

Let   $T_i$ ($i=1,2,\cdots, L$) be the time taken to move from $v_{\tau_{i-1}}$ to $v_{\tau_{i}}$, namely $T_i=\tau_i-\tau_{i-1}$. Therefore the return time $T_{H_0\rightarrow H_0}(t)$ on $G(t)$ satisfies
\begin{eqnarray}
\label{pathlength}
  T_{H_0\rightarrow H_0}(t)&=& \tau_L - \tau_0+T^a_{C\rightarrow C}(t-1) \nonumber \\
   &=& T_1+T_2+\cdots+T_L+T^a_{H_0\rightarrow H_0}(t).
\end{eqnarray}

Because $v_{\tau_i}=H_0$ (or $H_u$) for any $i=0, 1,2,\cdots$, $L$ and $v_{\tau_i} \neq v_{\tau_{i-1}}$ for any $i>0$. Then $T_i$ ($i=1,2,\cdots$, $L$) are identically distributed random variables, each of them is the first-passage time from  hub $H_0$ to  hub $H_u$ (or from hub $H_u$ to  hub $H_0$). Its  probability generating function is denoted by $ \Phi_{FPT}(t,z)$.

Note that  $L$, $T^a_{H_0\rightarrow H_0}(t)$, $T_1$, $T_2$, $\cdots$  are independent random variables. According to the properties (see Eqs.~(\ref{Sum_PGF2}) and (\ref{Sum_PGFn})) of the  probability generating function presented in Appendix~\ref{sec:PGF}, the probability generating function of return time $T_{H_0\rightarrow H_0}(t)$  satisfies
\begin{equation}\label{RR_PGF_RTO}
  \Phi_{RT}(t,z)=\left. \Phi_{RT}(0,x)\right|_{x=\Phi_{FPT}(t,z)}*\Phi^a_{RT}(t,z).
\end{equation}
As derived in Appendix~\ref{PGF_aRTt}, for any $t>0$,
\begin{equation}\label{Rec_PGF_RT_a}  
  \Phi^a_{RT}(t,z)=\Phi^a_{RT}(1,\Phi_{FPT}(t-1,z))*\Phi^a_{RT}(t-1,z).
\end{equation}

Replacing $\Phi^a_{RT}(t,z)$ from Eq.~ (\ref{Rec_PGF_RT_a}) in Eq.~ (\ref{RR_PGF_RTO}),
for any $t>0$,
 \begin{eqnarray}\label{DR_PGF_RT}
 &&\Phi_{RT}(t,z)    \nonumber \\
 &=&\Phi_{RT}(0,\Phi_{FPT}(t,z))* \Phi^a_{RT}(1,\Phi_{FPT}(t\!-\!1,z))*\Phi^a_{RT}(t\!-\!1,z)        \nonumber \\
     &=& \frac{\Phi_{RT}(0,\Phi_{FPT}(t,z))* \Phi^a_{RT}(1,\Phi_{FPT}(t\!-\!1,z))}{\Phi_{RT}(0,\Phi_{FPT}(t\!-\!1,z))} \nonumber \\
     & & *\Phi_{RT}(0,\Phi_{FPT}(t\!-\!1,z))* \Phi^a_{RT}(t\!-\!1,z)\nonumber \\
     &=&\frac{\Phi_{RT}(0,\Phi_{FPT}(t,z))* \Phi^a_{RT}(1,\Phi_{FPT}(t\!-\!1,z))}{\Phi_{RT}(0,\Phi_{FPT}(t\!-\!1,z))} \nonumber \\
     & & *\Phi_{RT}(t\!-\!1,z).
\end{eqnarray}
As derived in Appendix~\ref{PGF_FPTt}, $$\Phi_{FPT}(t,z)=\Phi_{FPT}(1,\Phi_{FPT}(t-1,z)).$$
Therefore,
 \begin{eqnarray}
 &&\Phi_{RT}(0,\Phi_{FPT}(t,z))  \nonumber \\
 &=&\Phi_{RT}(0,\Phi_{FPT}(1,\Phi_{FPT}(t-1,z))) \nonumber \\
     &=&\left. \Phi_{RT}(0,\Phi_{FPT}(1,x))\right|_{x=\Phi_{FPT}(t-1,z)}.
 \end{eqnarray}
 Thus
 \begin{eqnarray}
 &&\Phi_{RT}(0,\Phi_{FPT}(t,z))* \Phi^a_{RT}(1,\Phi_{FPT}(t\!-\!1,z))  \nonumber \\
     &=&\left. \left[\Phi_{RT}(0,\Phi_{FPT}(1,x))*\Phi^a_{RT}(1,x)\right]\right|_{x=\Phi_{FPT}(t-1,z)} \nonumber \\
     &=& \left. \Phi_{RT}(1,x)\right|_{x=\Phi_{FPT}(t-1,z)}.
 \end{eqnarray}
Let
 $$\psi(z)\equiv\frac{\Phi_{RT}(0,z)}{\Phi_{RT}(1,z)}.$$
We have
 \begin{eqnarray}
&&\frac{\Phi_{RT}(0,\Phi_{FPT}(t,z))* \Phi^a_{RT}(0,\Phi_{FPT}(t\!-\!1,z))}{\Phi_{RT}(0,\Phi_{FPT}(t\!-\!1,z))} \nonumber \\
&=&\frac{1}{\left. \psi(x)\right|_{x=\Phi_{FPT}(t-1,z)}}.
\end{eqnarray}
Inserting it into Eq.~ (\ref{DR_PGF_RT}), we obtain Eq.~(\ref{R_RT}).

Because the network of generation $0$ is just two nodes connected by an edge, the return probability of return in odd times is $0$ and the return probability of return in even times is $1$. Therefore
 \begin{eqnarray}
 \Phi_{RT}(0,z)=\sum_{n=0}^{+\infty}z^{2n}=\frac{1}{1-z^2}.
 \end{eqnarray}

\section{Derivation of Eq.~(\ref{Rec_PGF_RT_a})}
 \label{PGF_aRTt}
 Considering any return path $\pi$  of $H_0$ on $G(t)$ in the presence of an absorbing hub $H_u$. It can be written as $$\pi=(v_0=H_0,v_1,v_2\cdots,v_{T^a_{H_0\rightarrow H_0}(t)}=H_0),$$
 where $v_i$ is the node  reached at time $i$ and $T^a_{H_0\rightarrow H_0}(t)$ is the length of the path $\pi$.
Let  $\Omega\equiv\{H_0,H_1,\cdots, H_{w-1}\}$. Similar to Appendix~\ref{sec:Rec_relation}, we introduce the observable $\tau_i=\tau_i(\pi)$,
which is defined recursively as Eqs.~(\ref{OB1}) and (\ref{OB2}).
Then,   the path $\pi$ can be restated as a ``simplified path'' defined as
\begin{equation}
\label{Def_simp}
  \iota(\pi)=(v_{\tau_{0}}=H_0,v_{\tau_1},\cdots, v_{\tau_L}=H_0 ),
\end{equation}
where $v_{\tau_{0}}\in \Omega$ and $L$ is the total number of observables obtained from the path $\pi$.

For any return path  $\pi$ of $H_0$ in the presence of an absorbing hub $H_u$, similar to the discussion in  Appendix~\ref{sec:Rec_relation},  path  $\pi$ includes a sub-path from $H_0$ to $H_0$ after node $v_{\tau_L}$. The sub-path does not reach any  node in  $\Omega$ except for $H_0$.  Therefore, the sub-path can be regarded as a return path of $H_0$ on $G(t)$ in the presence of $w-1$ absorbing hubs (i.e., $H_1$,  $H_2$, $\cdots$,   $H_{w-1}$). In fact, the sub-path only includes nodes of  $\Gamma_1$ and $\Gamma_w$ (see Fig.~\ref{Self_similar}), which are  copies of $G(t-1)$. By symmetry,  nodes of $\Gamma_1$ are in one to one corresponding with nodes of $\Gamma_w$. If we replaced all the nodes of  $\Gamma_w$ with the corresponding nodes of $\Gamma_1$ in the sub-path, we obtain a return path of $H_0$ in $\Gamma_1$ which never reaches hub $H_1$. It is a return path of $H_0$ on $\Gamma_1$ in the presence of an absorbing hub  $H_1$  and has the same path length as the original sub-path. If we look $\Gamma_1$ as a copy of $G(t-1)$, hub $H_1$ of $G(t)$ can also be looked as $H_u$ of $G(t-1)$.   Therefore the  length of the sub-path  after node $v_{\tau_L}$ can be regarded as the return time of $H_0$ on $G(t-1)$ in the presence of an absorbing hub  $H_u$.

Let  $T_i=\tau_i-\tau_{i-1}$ for any $i=1,2,\cdots, L$. Therefore the length of the path $\pi$ satisfies
\begin{equation}
\label{RTa_pathlength}
  T^a_{H_0\rightarrow H_0}(t)= T_1+T_2+\cdots+T_L+ T^a_{H_0\rightarrow H_0}(t-1).
\end{equation}

Note that $\Omega$ just represents  the nodes set of the (u,v) flower with generation $1$ and the path $\iota(\pi)$ includes only the  nodes of $\Omega$. Thus, $\iota(\pi)$ is just a return path of $H_0$  on $G(1)$ in the presence of an absorbing hub $H_u$ and $L$ is just the path length. Therefore, the probability generating function of $L$ is $ \Phi^a_{RT}(1,z)$.

Similarly, we find that $T_i$ ($i=1,2,\cdots$) are identically distributed random variables, each of them can be regarded as the first-passage time from  hub $H_0$ to $H_u$  on $G(t-1)$. Therefore  the probability generating function of $T_i$ ($i=1,2,\cdots$) is $\Phi_{FPT}(t-1,z)$. Thus, we can obtain Eq.~(\ref{Rec_PGF_RT_a}) from Eqs.~(\ref{Sum_PGF2}), (\ref{Sum_PGFn}) and (\ref{RTa_pathlength}).


\section{Probability generating function of FPT from $H_0$ to $H_u$}
 \label{PGF_FPTt}
For the $(u$, $v)$ flower of generation $t=1$, the probability generating function of  FPT from $H_0$ to $H_u$ is presented in Eq.~(\ref{PhiFPT0}). 
For the $(u$, $v)$ flower of generation $t>1$, let $T_{H_0\rightarrow H_u}(t)$ denotes the FPT $H_0$ to $H_u$. Similar to the derivation of Eq.~(\ref{RTa_pathlength}), we can find independent random variables $L$,  $T_1$, $T_2$, $\cdots$, such that
\begin{equation}
\label{FPT_pathlength}
  T_{H_0\rightarrow H_u}(t)= T_1+T_2+\cdots+T_L.
\end{equation}
Here $L$ is the first-passage time from hub $H_0$ to $H_u$  on $G(1)$ and $T_i$ ($i=1,2,\cdots$) are identically distributed random variables, each of them is the first-passage time from  hub $H_0$ to $H_u$ on $G(t-1)$. Therefore the probability generating function of $L$ is $\Phi_{FPT}(1,z)$ and the probability generating function of $T_i$ ($i=1,2,\cdots$) is $\Phi_{FPT}(t-1,z)$. Thus, we can obtain from Eqs.~(\ref{Sum_PGF2}), (\ref{Sum_PGFn}) and (\ref{FPT_pathlength}) that, for any $t>1$,
\begin{equation}\label{Rec_PGF_FPT}
  \Phi_{FPT}(t,z)=\Phi_{FPT}(1,\Phi_{FPT}(t-1,z)).
\end{equation}
By taking the first order derivative on both sides of Eq.~(\ref{Rec_PGF_FPT}) and posing $z=1$,
we obtain the mean FPT from $H_0$ to $H_u$, i.e., for any $t\geq1$,
\begin{eqnarray}\label{MFPT}
    \langle FPT_t \rangle&=&\left. \frac{\partial}{\partial z}\Phi_{FPT}(t,z)\right|_{z=1}\nonumber \\
                 &=& \left. \frac{\partial}{\partial z}\Phi_{FPT}(1,z)\right|_{z=1}\times\left. \frac{\partial}{\partial z}\Phi_{FPT}(t-1,z)\right|_{z=1}\nonumber \\
                 &=&\langle FPT_1 \rangle \langle FPT_{t-1} \rangle               \nonumber \\
                 &=&\cdots=\langle FPT_1 \rangle^{t}.
\end{eqnarray}

For $t=1$, the $(u,v)$ flower $G(1)$ is a ring with $w$ nodes and $w$ links. If we view the networks $G(1)$ as electrical networks  by considering each edge to be a unit resistor, the effective resistance  between two nodes $H_0$ and $H_u$ is $R_{H_0\leftrightarrow H_u}=\frac{uv}{u+v}$. Therefore, the mean FPT from $H_0$ to $H_u$ is~\cite{Te91}
\begin{equation}
\langle FPT_1 \rangle=E_0R_{H_0\leftrightarrow H_u}=uv.
\label{F1}
\end{equation}
Inserting Eq.~(\ref{F1}) into Eq.~(\ref{MFPT}), for any $t\geq1$,
\begin{eqnarray}\label{MFPT_E}
    \langle FPT_t \rangle=(uv)^{t}.
\end{eqnarray}
It is easy to verify that Eq.~(\ref{MFPT_E}) also hold for $t=0$.
Since the volume of the underlying structure scales as $N_t \sim (u+v)^t$ for large sizes, the previous expression shows
\begin{eqnarray}\label{MFPT_S}
    \langle FPT_t \rangle \sim N_t^\frac{ln(uv)}{ln(u+v)}.
\end{eqnarray}

Similarity, we can also obtain the second moment of the FPT from $H_0$ to $H_u$, referred to as $\langle FPT^2_t\rangle$. Let $\Theta_t\equiv\left. \frac{\partial^2}{\partial z^2}\Phi_{FPT}(t,z)\right|_{z=1}$.
By taking the second order derivative on both sides of Eq.~(\ref{Rec_PGF_FPT}) and posing $z=1$, for any $t\geq1$,
\begin{eqnarray}\label{Rec_d2FPT}
   \Theta_t&=&\langle FPT_1 \rangle\Theta_{t-1}+\Theta_{1}\left(\langle FPT_{t-1} \rangle\right)^2\nonumber \\
                 &=& uv\Theta_{t-1}+\Theta_{1}(uv)^{2t-2}\nonumber \\
                 &=&(uv)^2\Theta_{t-2}+\Theta_{1}(uv)^{2t-3}+\Theta_{1}(uv)^{2t-2}\nonumber \\
                 &=&\cdots             \nonumber \\
                 &=&(uv)^{t-1}\Theta_{1}+\Theta_{1}\left[(uv)^{t}+(uv)^{t+1}+\cdots+\Theta_{1}(uv)^{2t-2}\right]\nonumber \\
                 &=&\Theta_{1}(uv)^{t-1}\frac{(uv)^{t}-1}{uv-1}.
\end{eqnarray}
For $\Theta_1$, it can be calculated directed by calculating the second order derivative of $\Phi_{FPT}(1,z)$ and the method for calculating $\Phi_{FPT}(1,z)$ is present in Appendix~\ref{sec:PGF}.
Therefore, for any $t\geq1$, 
\begin{eqnarray}\label{M2FPT}
   \langle FPT^2_t\rangle&=&\Theta_t+\langle FPT_t \rangle\nonumber \\
                 &=&\Theta_{1}(uv)^{t-1}\frac{(uv)^{t}-1}{uv-1}+(uv)^{t},
\end{eqnarray}
where $\Theta_{1}$ can also be rewritten as $\langle FPT^2_1\rangle\!-\!\langle FPT_1\rangle$.

It is easy to obtain $ \langle FPT^2_0\rangle=1$. Therefore Eq.~(\ref{M2FPT})  also holds for $t=0$.

\end{document}